\documentclass[sigconf,authorversion]{acmart}

\usepackage{subcaption}

\AtBeginDocument{%
  }

\setcopyright{acmlicensed}
\copyrightyear{2026}
\acmYear{2026}
\acmDOI{XXXXXXX.XXXXXXX}

\acmConference[AlpCHI '26]{AlpCHI 2026}{March 01--05,
  2026}{Ascona, Switzerland}

\acmISBN{978-1-4503-XXXX-X/2018/06}

\begin{document}

\title[Heads Up!: Photogrammetry Annotations and Augmented Reality Visualizations for Guided Backcountry Skiing]{Heads Up!: Towards In Situ Photogrammetry Annotations and Augmented Reality Visualizations for Guided Backcountry Skiing}

\author{Christoph Albert Johns}
\affiliation{%
  \institution{University of Oldenburg}
  \city{Oldenburg}
  \country{Germany}
}
\email{christoph.johns@uni-oldenburg.de}

\author{László Kopácsi}
\affiliation{%
  \institution{German Research Center for Artificial Intelligence (DFKI)}
  \city{Saarbrücken}
  \country{Germany}
}
\email{laszlo.kopacsi@dfki.de}

\author{Michael Barz}
\affiliation{%
  \institution{German Research Center for Artificial Intelligence (DFKI)}
  \city{Saarbrücken}
  \country{Germany}
}
\email{michael.barz@dfki.de}

\author{Daniel Sonntag}
\affiliation{%
  \institution{University of Oldenburg}
  \city{Oldenburg}
  \country{Germany}
}
\affiliation{%
  \institution{German Research Center for Artificial Intelligence (DFKI)}
  \city{Saarbrücken}
  \country{Germany}
}
\email{daniel.sonntag@dfki.de}


\begin{abstract}
Backcountry skiing is an activity where a group of skiers navigate challenging environmental conditions to ski outside of managed areas.
This activity requires careful monitoring and effective communication around the current weather and terrain conditions to ensure skier safety.
We aim to support and facilitate this communication by providing backcountry guides with a set of \textit{in situ} spatial annotation tools to communicate hazards and appropriate speeds to the ski recreationalists.
A guide can use a tablet application to annotate a photogrammetry-based map of a mountainside, for example, one collected using a commercial camera drone, with hazard points, slow-down zones, and safe zones.
These annotations are communicated to the skiers via visual overlays in augmented reality heads-up displays.
We present a prototype consisting of a web application and a virtual reality display that mirror the guide's and skier’s perspectives, enabling participatory interaction design studies in a safe environment.

\end{abstract}

\ccsdesc[500]{Human-centered computing~Interactive systems and tools}

\keywords{Backcountry skiing, augmented reality, heads-up display, map interfaces, drones, UAVs}

\begin{teaserfigure}
  \includegraphics[width=\linewidth]{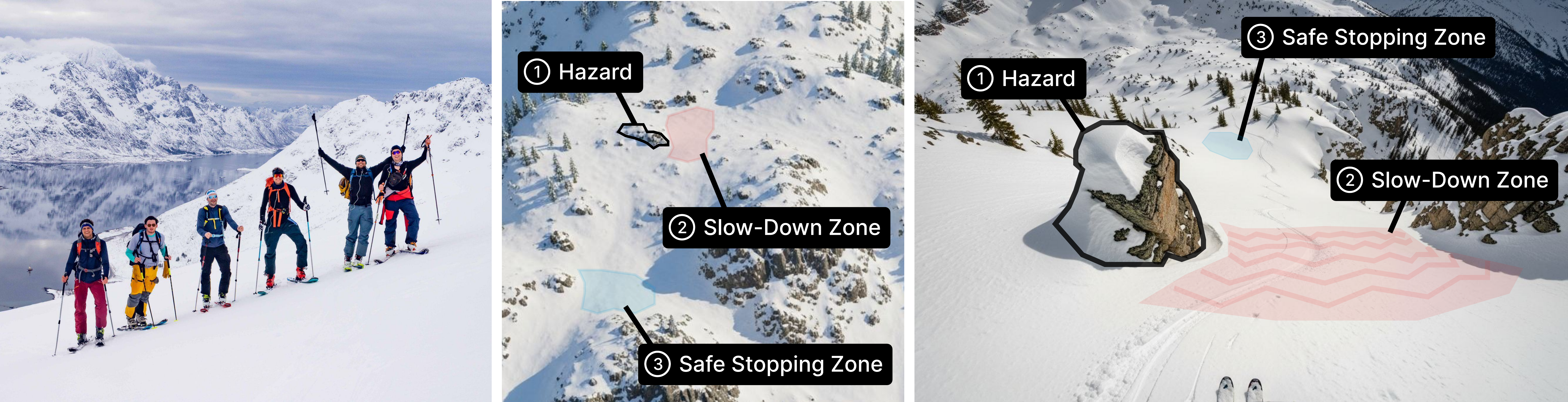}
  \caption{Backcountry skiing is an activity where a group of skiers scale a mountainside to ski natural slopes outside of managed areas. It requires clear communication between a professional guide and the recreational skiers to stay safe, minimize environmental impact, and remain exciting and enjoyable. Left: A group of skiers on a guided backcountry skiing tour in Norway. Center: Concept for an annotated photogrammetry map of a mountainside slope. The 3D map displays hazards, slow-down zones, and safe zones authored by the guide to communicate safety risks to the recreational skiers. Right: The skiers view the annotations via augmented reality overlays in a heads-up display. In our demo, participants act out the guide-to-skier communication using a web application and a virtual reality setup. Image Source: Edouard de Becker Remy (left), Google Imagen 4 (center, right) with custom annotations.}
  \Description{A composite figure with three images. On the left, a group of off-piste skiers stands on a snowy mountain. In the center, a top-down image of a mountainside is annotated with a hazard marker, a slow-down zone, and a safe zone. On the right, a first-person image from the skiers point of view shows a descending slope with a rock outlined and labeled "hazard", an area close to the rock highlighted in red and labeled "slow-down zone", and an area far down the slope highlighted in blue and labeled "safe zone".}
  \label{fig:teaser}
\end{teaserfigure}

\maketitle

\section{Introduction}

\begin{figure}[tbp]
    \centering
    \includegraphics[width=\linewidth,trim={0 2cm 0 2cm},clip]{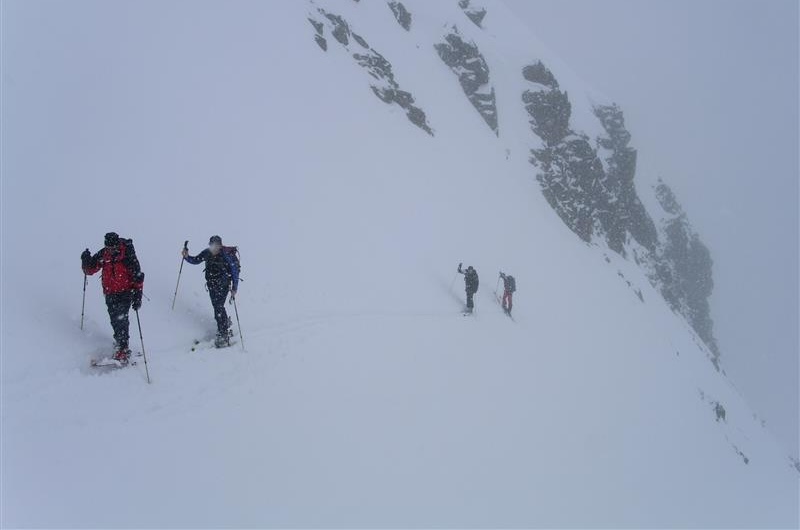}
    \caption{Backcountry skiing presents inherent dangers due to the uncertain terrain, risk of avalanches, and risk of low visibility. It thus requires effective communication among the group members to prevent harm to people and the environment. Image Source: Michael Barz}
    \Description{A group of skiers navigating a rocky mountain cliff in the fog.}
    \label{fig:mountain}
\end{figure}

Backcountry skiing (closely related to \textit{off-piste} skiing, \textit{freeriding}, and \textit{ski mountaineering}) is an activity where a group of skiers scale a mountainside to ski natural slopes outside of managed areas.
This often presents inherent dangers, such as uncertain terrain and natural obstacles (\autoref{fig:mountain}).
Navigating these challenges requires clear communication between the guide and skiers to stay safe, minimize environmental impact, and ensure an exciting and enjoyable experience \cite{fedosov_start_2015}.
Technology plays an important role in enabling and augmenting this communication.

Despite recognizing the need to support risk mitigation in backcountry skiing \cite{wozniak_soil_2017}, research to facilitate communication in skiing has so far primarily focused on on-piste environments (e.g., around performance or social activities \cite{fedosov_skiar_2016, fedosov_design_2016, niforatos_augmenting_2017}) or recovery from dangers (e.g., in avalanche search and rescue \cite{desjardins_collaboration_2014, rahman_collab-sar_2018, stasik_autonomous_2025}).
We propose a system to prevent harm in hazardous off-piste environments by supporting communication between backcountry skiing guides and recreational skiers regarding safety issues, utilizing \textit{in situ} photogrammetry annotations and visual overlays in augmented reality (AR) heads-up displays (HUDs).

Our system concept leverages technology already carried on many backcountry tours, such as tablet computers and commercial camera drones, and anticipates future technology (AR ski goggles) to address monitoring and communication needs where guides are typically cut off from communication with recreational skiers: after 'dropping into' a slope section.
At the same time, the system allows guides to regulate the amount of information communicated to skiers, preserving the sensory experience of the most rewarding aspect of backcountry skiing: the descent \cite{wozniak_soil_2017}.

Our prototype implements this concept using a web application and a virtual reality setup displaying a 3D model of a mountainside slope (\autoref{fig:web_mockup}).
A participant, acting as a guide, can annotate the virtual environment using the web application running on a tablet computer with hazard, slow-down, and safe stopping zones.
Annotations can be previewed by the same or another participant in a first-person view using the web application or a head-mounted display (HMD), such as the HTC Vive Focus Vision.
The HMD displays the skier's point of view, including the guide's annotations, from a position on the slope.
The system can also be used by two participants simultaneously, with one acting as a guide using the web application and the other as a recreational skier wearing the HMD, communicating potential safety concerns via spatial annotations while navigating down the slope.

Our demo aims to provide insights into effective guide-to-recre-ationalist communication in extreme natural environments.
It prepares a later evaluative study with broader implications for communication strategies to prevent harm to recreationalists and the environment in areas such as hiking, climbing, or mountain biking \cite{wozniak_soil_2017}.

In summary, we present the following demo:

\begin{itemize}
    \item A multi-device system, consisting of a tablet computer and an HMD (HTC Vive Focus Vision), that simulates an \textit{in situ} \textbf{photogrammetry annotation} and display tool \textbf{for guided backcountry skiing}.
    \item The \textbf{tablet} computer displays a 3D map of an off-piste mountainside slope. A user, playing the role of a backcountry guide, can \textbf{annotate} this map \textbf{with hazards, slow-down, and safe zones} using touch or stylus interaction.
    \item The \textbf{HMD} shows the skier's perspective. A user, playing the role of a recreational skier, can \textbf{view} the guide's spatial \textbf{annotations as anchored overlays} that direct the skier's attention to hazards, areas where speed should be reduced, and safe stopping areas to meet up with other group members, giving insights into the effectiveness of the annotations.
\end{itemize}

\section{Related Work}

We envision a system that includes an AR HUD for safety-related guide-to-recreationalist communication and an \textit{in situ} terrain-map-ping-and-annotation system centered around a tablet computer and a commercial camera drone.
It applies them to the domain of backcountry skiing.
We therefore review how skiing has been examined in human-computer interaction (HCI) research, how HUDs have been applied in outdoor and safety contexts, and how \textit{in situ} terrain mapping and annotation have been applied in mountainside settings.

\subsection{Skiing in Human-Computer Interaction}

HCI research has primarily aimed to support skiing activities through the lenses of skill assessment and improvement and social communication.
It has mainly focused on on-piste and simulator settings.
For example, Zhang \textit{et al.}~\cite{zhang_watch-your-skiing_2021} and Tabei \textit{et al.}~\cite{tabei_skiexargame_2025} developed virtual and augmented reality systems for posture training in simulated skiing environments.
Similarly, Hoffard \textit{et al.}~\cite{hoffard_pushtoski_2021} used a virtual reality setup to improve skiing posture through haptic feedback.

For multi-person communication in skiing contexts, HCI research has primarily aimed at improving social connectedness, for example, through supporting the sharing of location-tagged photos, videos, points of interest, and hazards in managed ski areas \cite{fedosov_skiar_2016}, or at improving accessibility in skiing through supporting guide-to-skier communications.
For example, several works have focused on providing guidance cues to blind skiers, for instance, through haptics \cite{aggravi_haptic_2016, haladjian_vihapp_2017} or audio feedback \cite{hirano_slopenav_2025}.
Most recently, Motahar \textit{et al.}~\cite{motahar_understanding_2025} investigated guidance in the context of competitive skiers with tetraplegia, finding that collaboration and safety, but also self-efficacy and independence, are core themes in assisted skiing.

Our system builds on these lines of research and applies them to safety-related guide-to-skier communication in unmanaged off-piste environments.
Its primary goal is to prevent harm to skiers and the environment.
Our work can best be viewed as a follow-up to Fedosov \textit{et al.}'s~\cite{fedosov_start_2015, fedosov_towards_2016} participatory design research where backcountry skiers were observed, interviewed, and co-designed technology to support communication in off-piste contexts, including a concept for an AR ski goggle application to support off-piste safety communication \cite{fedosov_start_2015}.

\begin{figure}[tbp]
    \centering
    \includegraphics[width=\linewidth,trim={0 7cm 0 9cm},clip]{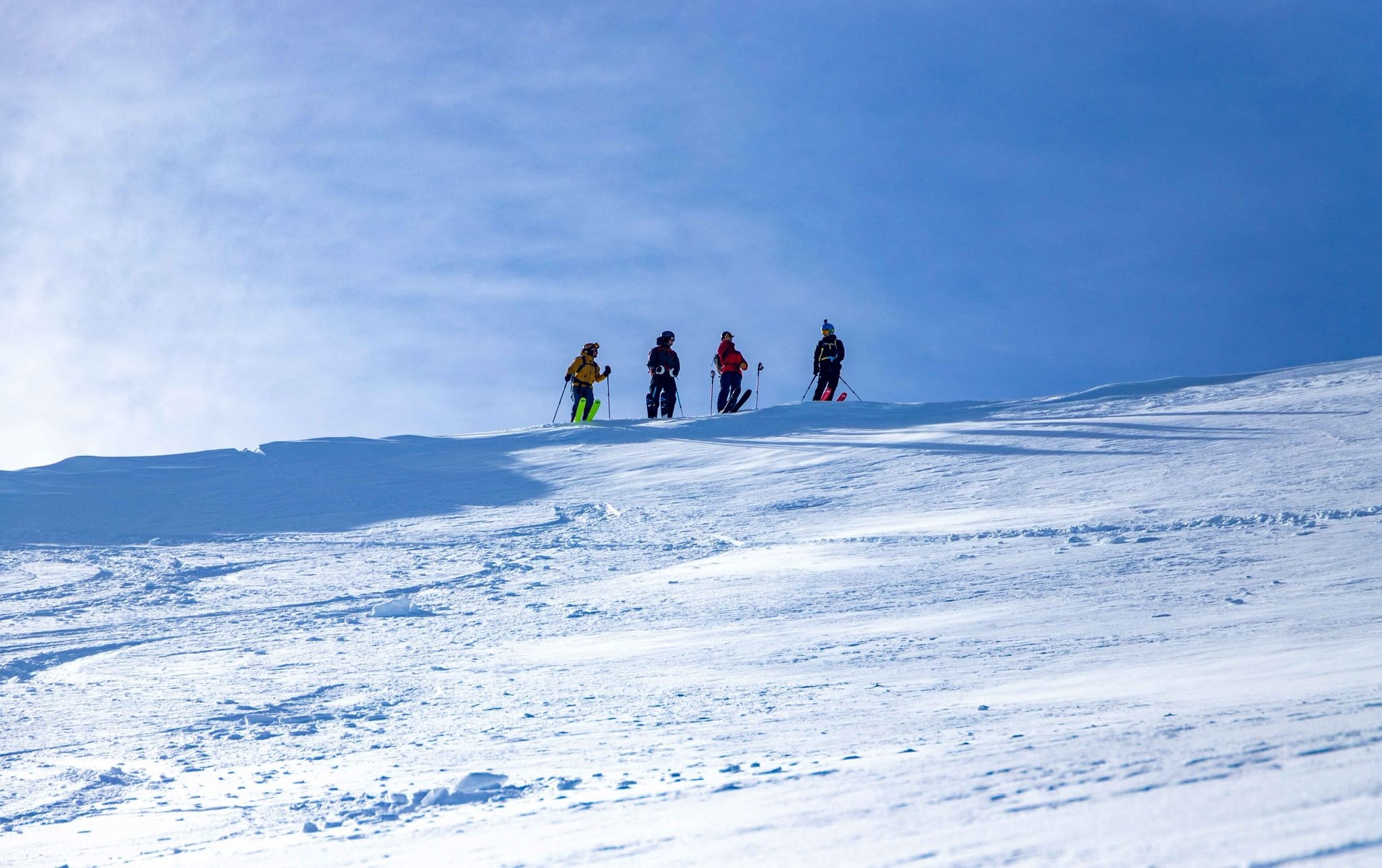}
    \caption{Before dropping in, the guide communicates with the group of skiers about the potential hazards and offers recommendations about how to approach the next section, commonly via a walkie-talkie. Image Source: Edouard de Becker Remy}
    \Description{A group of freeride skiers on the edge of a cliff, waiting to drop in.}
    \label{fig:before_drop}
\end{figure}

\subsection{Heads-Up Displays in Outdoor and Safety Contexts}

HUDs have been shown to be effective in communicating safety-critical information, including through the use of AR features.
This has primarily been demonstrated in applications related to assisted driving \cite{goodge_can_2024, wang_arive_2025}.
HUDs have also been used in the context of outdoor activities and outdoor sports, such as skiing \cite{niforatos_augmenting_2017, niforatos_s-helmet_2016} or cycling \cite{he_pedalling_2024, von_sawitzky_how_2024, zhao_hazardsnap_2024, huang_safety_2025}.
These works mainly aim to increase users' awareness of their surroundings, particularly regarding hazards.
At the same time, HUD information must avoid diverting the user's attention from the primary activity, a phenomenon known as inattentional blindness \cite{wang_inattentional_2022}.

For example, several works have investigated HUDs to improve safety in cycling, with applications ranging from navigation \cite{he_pedalling_2024} to hazard notifications \cite{von_sawitzky_how_2024, zhao_hazardsnap_2024} and safe overtaking \cite{huang_safety_2025}.
In the context of skiing, research has investigated how skiers' peripheral vision can be augmented with information about skiers coming from behind to prevent on-piste accidents \cite{niforatos_augmenting_2017, niforatos_s-helmet_2016} and speculated how it may help inform skiers of potential hazards and safe paths in off-piste contexts \cite{fedosov_start_2015}.

Together, work on HUDs in safety-critical environments has emphasized that the intent to prevent harm through notifications and annotations must be balanced with minimizing the amount of visual information displayed to the user to limit their overall cognitive load and excessive diversion of attention.


\subsection{In Situ Mountainside Terrain Mapping and Annotation}

Terrain mapping in the context of on-piste and backcountry skiing has primarily focused on preparatory route planning and \textit{in situ} search-and-rescue situations, generally aiming at collaborative and social contexts.
For example, Wiehr \textit{et al.}~\cite{wiehr_artopos_2017} created a mobile AR system to visualize and collaboratively annotate a topographical mountainside map with routes for ski, hiking, or climbing trips.
Similarly, Fedosov \textit{et al.}~\cite{fedosov_skiar_2016,fedosov_design_2016} developed an AR system to annotate ski resort maps with location-anchored content, such as photos, routes, or hazard markers.

In the context of backcountry and off-piste mountainside environments, \textit{in situ} terrain mapping and annotation have been used to effectively conduct and communicate during avalanche search-and-rescue missions \cite{rahman_collab-sar_2018}, including using lightweight drones to map surroundings \cite{rahman_collab-sar_2018} and locate avalanche transceiver devices \cite{stasik_autonomous_2025}.
These works have emphasized the importance of flexible devices and collaborative features that facilitate and enrich communication while remaining lightweight and durable in the challenging outdoor conditions.

\begin{figure}[b]
    \centering
    \includegraphics[width=\linewidth]{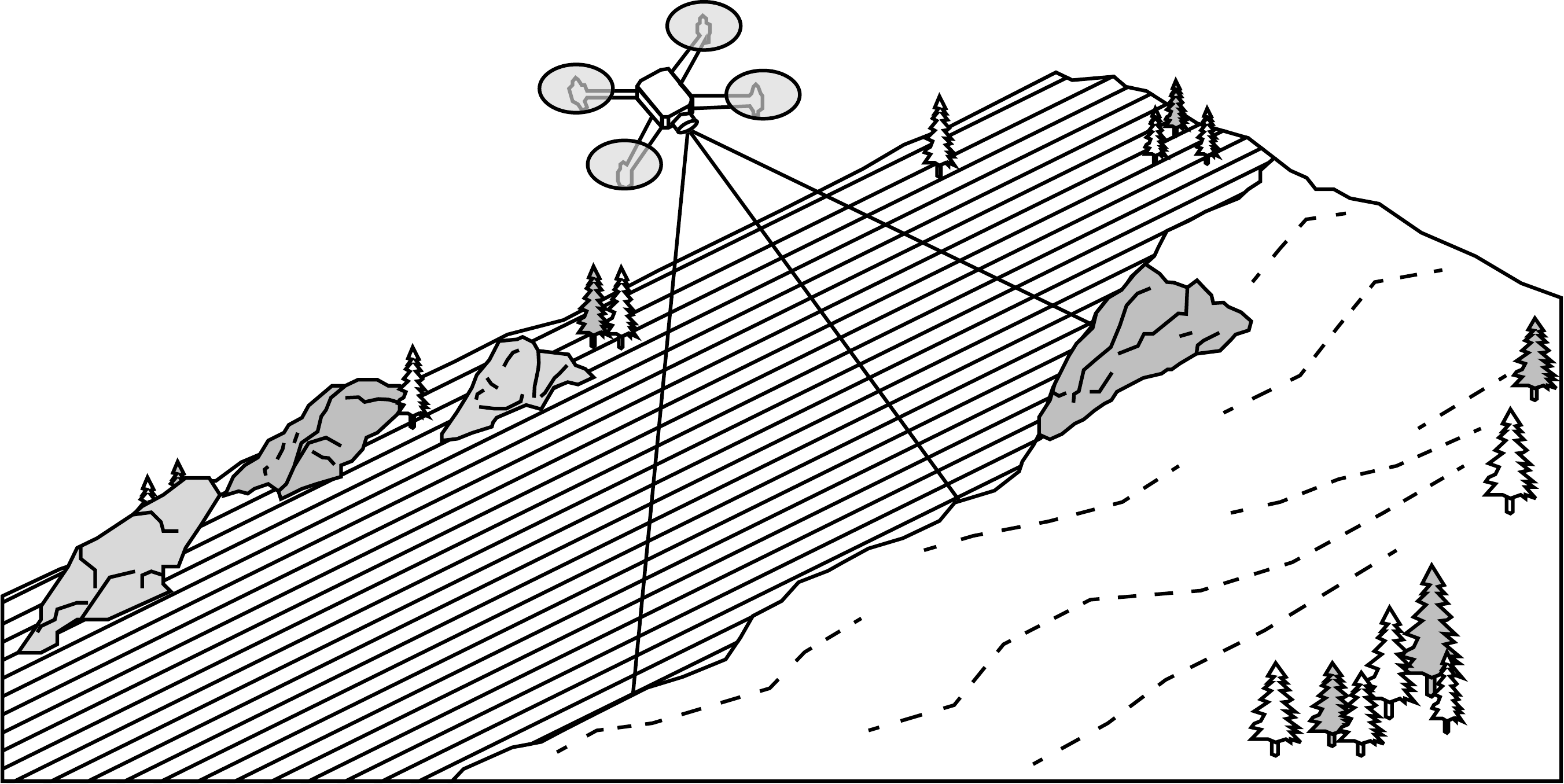}
    \caption{A commercial camera drone flies up to map the terrain and conditions \textit{in situ}. Using the generated 3D map, the guide can annotate hazards, appropriate speeds, and safe stopping zones to communicate to the skiers.}
    \Description{A drone mapping a mountainside.}
    \label{fig:mapping}
\end{figure}

\begin{figure*}[htbp]
    \centering
    \includegraphics[width=.9\linewidth]{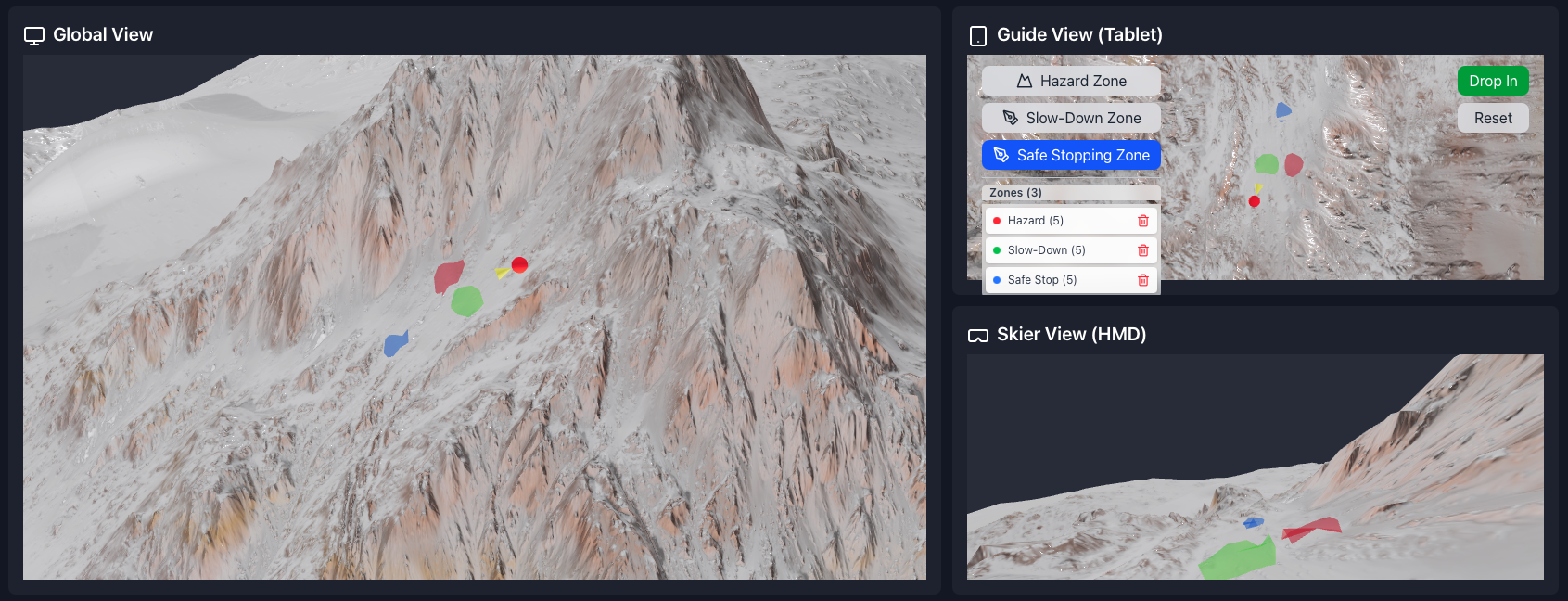}
    \caption{We demonstrate the envisioned multi-device application using a multi-view 3D web application and a commercial head-mounted display (HMD). The user playing the guide can add hazards and draw zones onto a 3D model of a mountainside slope and preview the skier's perspective using the web application or using the HMD.}
    \Description{A 3D web application displaying a global view of a virtual plane, an orthographic top-down view of the same plane, and a first-person view from a position on the surface of the plane.}
    \label{fig:web_mockup}
\end{figure*}

\section{System}

Backcountry skiing is usually conducted in groups led by a professional guide who plans an appropriate route together with the skiers, inspects the mountain on the walk up (during 'skinning'; \autoref{fig:mountain}), and continuously monitors weather and terrain conditions, as well as group dynamics.
During the descent, the guide ensures the safety and enjoyment of the skiers through continuous monitoring and communication (\autoref{fig:before_drop}).

We aim to support this guide-to-recreationalist communication through a multi-component system with two output devices operating in three interaction steps:
First, a lightweight commercial camera drone maps the mountain terrain (e.g., during ascent or before descent), providing the guide with a photogrammetry-based \textit{in situ} 3D map (\autoref{fig:mapping}).
The guide can then use a tablet application to annotate this map with hazards, slow-down, and safe stopping zones to draw the skiers' attention to safety-critical information (\autoref{fig:teaser}, center; \autoref{fig:web_mockup}).
The skiers receive the guide's annotations via AR HUDs, which display unobtrusive visual highlights.
These strive to draw minimal attention while ensuring the delivery of safety-critical information (\autoref{fig:teaser}, right).

We prototype this envisioned system by implementing a web application displayed on a tablet computer and a virtual reality application displayed on a commercial HMD (HTC Vive Focus Vision).
One user, in the role of the guide, can view and annotate a 3D model of a mountainside environment using the tablet computer running the web application.
They can make use of a free-to-navigate 3D view, a top-down view of the slope, and a preview of the skier's first-person perspective to set the annotations (\autoref{fig:web_mockup}).
Then, a user in the role of a recreational skier can view the annotations through the HMD from a position on the slope, mimicking the \textit{in situ} annotation-based guide-to-skier communication.
We aim to utilize this demo system in participatory interaction design studies to facilitate discussions with experienced guides and skiers regarding the communication of safety-critical information in backcountry contexts.


\section{Discussion}

Our system is designed to strengthen existing communication patterns between backcountry guides and recreational skiers, particularly concerning safety on unmanaged slopes.
The core tension addressed by this design lies between the recreational skier's need for minimal distraction to maximize the sensory enjoyment of the descent and the guide's need for control, monitoring, and ensuring safety.
The current system prioritizes the guide's autonomy and control, assuming the guide will provide only the minimum necessary hints to ensure safety while preserving the skier's enjoyment, an approach hinted at as acceptable in related work \cite{fedosov_start_2015}.
However, the principles explored here hold broader implications for other guided and hazardous outdoor activities, such as running, hiking, climbing \cite{wozniak_soil_2017}, cycling, or even more extreme environments like diving or cave exploration \cite{mencarini_becoming_2023}.

A significant challenge moving forward is extending the present prototype from a virtual reality environment to a safe and usable AR system in real-world conditions.
We anticipate that the demo will yield fruitful discussions with experienced tour guides and recreationalists in backcountry skiing or related disciplines, such as ski mountaineering. We aim to develop monitoring and communication technologies in close collaboration with guides and skiers, applying human-centered, participatory interaction design methodologies. Our demonstrator will also help investigate how safety features of existing map-based services, such as avalanche terrain hazard maps \cite{Outdooractive2025_ATHM}, can be effectively integrated into a backcountry guide's \textit{in situ} decision-making process.

The demo can also facilitate the evaluation of multimodal input technologies \cite{oviatt_handbook_2019,bittner_interactive_2025,barz_multisensor-pipeline_2021} based on a combination of, for example, gaze and speech.
These may enable more robust and secure communication in critical situations, such as steep slope environments where both hands may be occupied for stabilization (\autoref{fig:mountain}). 
Furthermore, the system could be extended to incorporate non-safety-related forms of communication common in backcountry skiing, such as mutual encouragement, or to merge with existing practices like recording drone and POV footage, enhancing its integration into the overall experience \cite{fedosov_start_2015,fedosov_design_2016}.

\section{Conclusion}

We propose a multi-device web and virtual reality system for exploring the interaction design of future \textit{in situ} photogrammetry mapping and annotations, supporting guide-to-recreationalist communication in backcountry skiing.
The demo aims to stimulate discussion on the effective and safe use of AR and HUD technology in extreme outdoor environments, as well as the development of effective communication tools for addressing safety-related issues within guided outdoor sports activities, particularly in the context of participatory interaction design studies.

\begin{acks}
This research is part of the Computational Sustainability \& Technology project area\footnote{\url{https://cst.dfki.de/}}, and has been supported by the Lower Saxony Ministry of Science and Culture (MWK) in the \textit{zukunft.niedersachsen} program, the Endowed Chair of Applied Artificial Intelligence at University of Oldenburg, and DFKI.
\end{acks}

\bibliographystyle{ACM-Reference-Format}
\bibliography{references,references_michael}

@inproceedings{motahar_understanding_2025,
	address = {New York, NY, USA},
	series = {{CHI} '25},
	title = {Understanding the {Training} {Experiences} of {Competitive} {Skiers} with {Tetraplegia}},
	isbn = {979-8-4007-1394-1},
	url = {https://dl.acm.org/doi/10.1145/3706598.3713627},
	doi = {10.1145/3706598.3713627},
	urldate = {2025-11-03},
	booktitle = {Proceedings of the 2025 {CHI} {Conference} on {Human} {Factors} in {Computing} {Systems}},
	publisher = {Association for Computing Machinery},
	author = {Motahar, Tamanna and Kim, YeonJae and Fisher, Eden and Wiese, Jason},
	month = apr,
	year = {2025},
	pages = {1--16},
}

@inproceedings{fedosov_skiar_2016,
	address = {New York, NY, USA},
	series = {{AH} '16},
	title = {{SkiAR}: {Wearable} {Augmented} {Reality} {System} for {Sharing} {Personalized} {Content} on {Ski} {Resort} {Maps}},
	isbn = {978-1-4503-3680-2},
	shorttitle = {{SkiAR}},
	url = {https://dl.acm.org/doi/10.1145/2875194.2875234},
	doi = {10.1145/2875194.2875234},
	urldate = {2025-11-04},
	booktitle = {Proceedings of the 7th {Augmented} {Human} {International} {Conference} 2016},
	publisher = {Association for Computing Machinery},
	author = {Fedosov, Anton and Elhart, Ivan and Niforatos, Evangelos and North, Alexander and Langheinrich, Marc},
	month = feb,
	year = {2016},
	pages = {1--2},
}

@inproceedings{tabei_skiexargame_2025,
	address = {New York, NY, USA},
	series = {{AHs} '25},
	title = {{SkiExARGame}: {Augmented} {Reality} {Exergame} for {Posture} {Correction} in {Alpine} {Ski}},
	isbn = {979-8-4007-1566-2},
	shorttitle = {{SkiExARGame}},
	url = {https://dl.acm.org/doi/10.1145/3745900.3747284},
	doi = {10.1145/3745900.3747284},
	urldate = {2025-11-04},
	booktitle = {Proceedings of the {Augmented} {Humans} {International} {Conference} 2025},
	publisher = {Association for Computing Machinery},
	author = {Tabei, Yuki and Matsumoto, Takashi and Peng, Yichen and Koike, Hideki},
	month = oct,
	year = {2025},
	pages = {467--469},
}

@inproceedings{fedosov_design_2016,
	address = {New York, NY, USA},
	series = {{MUM} '16},
	title = {Design and evaluation of a wearable {AR} system for sharing personalized content on ski resort maps},
	isbn = {978-1-4503-4860-7},
	url = {https://dl.acm.org/doi/10.1145/3012709.3012721},
	doi = {10.1145/3012709.3012721},
	urldate = {2025-11-04},
	booktitle = {Proceedings of the 15th {International} {Conference} on {Mobile} and {Ubiquitous} {Multimedia}},
	publisher = {Association for Computing Machinery},
	author = {Fedosov, Anton and Niforatos, Evangelos and Elhart, Ivan and Schneider, Teseo and Anisimov, Dmitry and Langheinrich, Marc},
	month = dec,
	year = {2016},
	pages = {141--152},
}

@inproceedings{fedosov_start_2015,
	address = {New York, NY, USA},
	series = {{MobileHCI} '15},
	title = {From {Start} to {Finish}: {Understanding} {Group} {Sharing} {Behavior} in a {Backcountry} {Skiing} {Community}},
	isbn = {978-1-4503-3653-6},
	shorttitle = {From {Start} to {Finish}},
	url = {https://dl.acm.org/doi/10.1145/2786567.2793698},
	doi = {10.1145/2786567.2793698},
	urldate = {2025-11-04},
	booktitle = {Proceedings of the 17th {International} {Conference} on {Human}-{Computer} {Interaction} with {Mobile} {Devices} and {Services} {Adjunct}},
	publisher = {Association for Computing Machinery},
	author = {Fedosov, Anton and Langheinrich, Marc},
	month = aug,
	year = {2015},
	pages = {758--765},
}

@inproceedings{aggravi_haptic_2016,
	address = {New York, NY, USA},
	series = {{AH} '16},
	title = {Haptic {Assistive} {Bracelets} for {Blind} {Skier} {Guidance}},
	isbn = {978-1-4503-3680-2},
	url = {https://dl.acm.org/doi/10.1145/2875194.2875249},
	doi = {10.1145/2875194.2875249},
	urldate = {2025-11-04},
	booktitle = {Proceedings of the 7th {Augmented} {Human} {International} {Conference} 2016},
	publisher = {Association for Computing Machinery},
	author = {Aggravi, Marco and Salvietti, Gionata and Prattichizzo, Domenico},
	month = feb,
	year = {2016},
	pages = {1--4},
}

@inproceedings{hirano_slopenav_2025,
	address = {New York, NY, USA},
	series = {{AHs} '25},
	title = {{SlopeNav}: {A} {Realtime} {Wearable} {Blind} {Ski} {Assistance} {System} with {Adaptive} {Path} {Planning} for {Simulated} {Environments}},
	isbn = {979-8-4007-1566-2},
	shorttitle = {{SlopeNav}},
	url = {https://dl.acm.org/doi/10.1145/3745900.3746082},
	doi = {10.1145/3745900.3746082},
	urldate = {2025-11-04},
	booktitle = {Proceedings of the {Augmented} {Humans} {International} {Conference} 2025},
	publisher = {Association for Computing Machinery},
	author = {Hirano, Toshihiro and Peng, Yichen and Kuribayashi, Masaki and Wu, Erwin and Morishima, Shigeo and Koike, Hideki},
	month = oct,
	year = {2025},
	pages = {369--380},
}

@inproceedings{desjardins_collaboration_2014,
	address = {New York, NY, USA},
	series = {{CSCW} '14},
	title = {Collaboration surrounding beacon use during companion avalanche rescue},
	isbn = {978-1-4503-2540-0},
	url = {https://dl.acm.org/doi/10.1145/2531602.2531684},
	doi = {10.1145/2531602.2531684},
	urldate = {2025-11-04},
	booktitle = {Proceedings of the 17th {ACM} conference on {Computer} supported cooperative work \& social computing},
	publisher = {Association for Computing Machinery},
	author = {Desjardins, Audrey and Neustaedter, Carman and Greenberg, Saul and Wakkary, Ron},
	month = feb,
	year = {2014},
	pages = {877--887},
}

@inproceedings{haladjian_vihapp_2017,
	address = {New York, NY, USA},
	series = {{UbiComp} '17},
	title = {{VIHapp}: a wearable system to support blind skiing},
	isbn = {978-1-4503-5190-4},
	shorttitle = {{VIHapp}},
	url = {https://dl.acm.org/doi/10.1145/3123024.3124443},
	doi = {10.1145/3123024.3124443},
	urldate = {2025-11-04},
	booktitle = {Proceedings of the 2017 {ACM} {International} {Joint} {Conference} on {Pervasive} and {Ubiquitous} {Computing} and {Proceedings} of the 2017 {ACM} {International} {Symposium} on {Wearable} {Computers}},
	publisher = {Association for Computing Machinery},
	author = {Haladjian, Juan and Reif, Maximilian and Brügge, Bernd},
	month = sep,
	year = {2017},
	pages = {1033--1037},
}

@inproceedings{niforatos_augmenting_2017,
	address = {New York, NY, USA},
	series = {{ISWC} '17},
	title = {Augmenting skiers' peripheral perception},
	isbn = {978-1-4503-5188-1},
	url = {https://dl.acm.org/doi/10.1145/3123021.3123052},
	doi = {10.1145/3123021.3123052},
	urldate = {2025-11-04},
	booktitle = {Proceedings of the 2017 {ACM} {International} {Symposium} on {Wearable} {Computers}},
	publisher = {Association for Computing Machinery},
	author = {Niforatos, Evangelos and Fedosov, Anton and Elhart, Ivan and Langheinrich, Marc},
	month = sep,
	year = {2017},
	pages = {114--121},
}

@inproceedings{wiehr_artopos_2017,
	address = {New York, NY, USA},
	series = {{UbiComp} '17},
	title = {{ARTopos}: augmented reality terrain map visualization for collaborative route planning},
	isbn = {978-1-4503-5190-4},
	shorttitle = {{ARTopos}},
	url = {https://dl.acm.org/doi/10.1145/3123024.3124446},
	doi = {10.1145/3123024.3124446},
	urldate = {2025-11-04},
	booktitle = {Proceedings of the 2017 {ACM} {International} {Joint} {Conference} on {Pervasive} and {Ubiquitous} {Computing} and {Proceedings} of the 2017 {ACM} {International} {Symposium} on {Wearable} {Computers}},
	publisher = {Association for Computing Machinery},
	author = {Wiehr, Frederik and Daiber, Florian and Kosmalla, Felix and Krüger, Antonio},
	month = sep,
	year = {2017},
	pages = {1047--1050},
}

@inproceedings{hoffard_pushtoski_2021,
	address = {New York, NY, USA},
	series = {{SIGGRAPH} '21},
	title = {{PushToSki} - {An} {Indoor} {Ski} {Training} {System} {Using} {Haptic} {Feedback}},
	isbn = {978-1-4503-8371-4},
	url = {https://dl.acm.org/doi/10.1145/3450618.3469158},
	doi = {10.1145/3450618.3469158},
	urldate = {2025-11-04},
	booktitle = {{ACM} {SIGGRAPH} 2021 {Posters}},
	publisher = {Association for Computing Machinery},
	author = {Hoffard, Jana and Nakamura, Takuto and Wu, Erwin and Koike, Hideki},
	month = aug,
	year = {2021},
	pages = {1--2},
}

@article{rahman_collab-sar_2018,
	title = {Collab-{SAR}: {A} {Collaborative} {Avalanche} {Search}-and-{Rescue} {Missions} {Exploiting} {Hostile} {Alpine} {Networks}},
	volume = {6},
	issn = {2169-3536},
	shorttitle = {Collab-{SAR}},
	url = {https://ieeexplore.ieee.org/document/8398223},
	doi = {10.1109/ACCESS.2018.2848366},
	urldate = {2025-11-04},
	journal = {IEEE Access},
	author = {Rahman, Md Arafatur and Azad, Saiful and Asyhari, A. Taufiq and Bhuiyan, Md Zakirul Alam and Anwar, Khoirul},
	year = {2018},
	pages = {42094--42107},
}

@article{stasik_autonomous_2025,
	title = {An {Autonomous} {Drone} for {Avalanche} {Search} and {Rescue}: {Integrating} {ArduPilot} and {Tracking}-{Beacon} {Detection}},
	volume = {13},
	issn = {2169-3536},
	shorttitle = {An {Autonomous} {Drone} for {Avalanche} {Search} and {Rescue}},
	url = {https://ieeexplore.ieee.org/document/11062844},
	doi = {10.1109/ACCESS.2025.3585245},
	urldate = {2025-11-04},
	journal = {IEEE Access},
	author = {Stasik, Jedrzej and Szandala, Tomasz},
	year = {2025},
	pages = {130758--130769},
}

@inproceedings{zhang_watch-your-skiing_2021,
	title = {Watch-{Your}-{Skiing}: {Visualizations} for {VR} {Skiing} using {Real}-time {Body} {Tracking}},
	shorttitle = {Watch-{Your}-{Skiing}},
	url = {https://ieeexplore.ieee.org/document/9585783},
	doi = {10.1109/ISMAR-Adjunct54149.2021.00088},
	urldate = {2025-11-04},
	booktitle = {2021 {IEEE} {International} {Symposium} on {Mixed} and {Augmented} {Reality} {Adjunct} ({ISMAR}-{Adjunct})},
	author = {Zhang, Xuan and Wu, Erwin and Koike, Hideki},
	month = oct,
	year = {2021},
	pages = {387--388},
}

@article{zhao_hazardsnap_2024,
	title = {{HazARdSnap}: {Gazed}-{Based} {Augmentation} {Delivery} for {Safe} {Information} {Access} {While} {Cycling}},
	volume = {30},
	issn = {1941-0506},
	shorttitle = {{HazARdSnap}},
	url = {https://ieeexplore.ieee.org/document/10319779},
	doi = {10.1109/TVCG.2023.3333336},
	number = {9},
	urldate = {2025-11-04},
	journal = {IEEE Transactions on Visualization and Computer Graphics},
	author = {Zhao, Guanghan and Orlosky, Jason and Gabbard, Joseph and Kiyokawa, Kiyoshi},
	month = sep,
	year = {2024},
	pages = {6378--6389},
}

@article{wang_arive_2025,
	title = {{ARive}: {Assisting} {Drivers} with {In}-{Car} {Augmented} {Reality} for {Risk} {Zone} {Detection}},
	volume = {9},
	shorttitle = {{ARive}},
	url = {https://dl.acm.org/doi/10.1145/3712270},
	doi = {10.1145/3712270},
	number = {1},
	urldate = {2025-11-04},
	journal = {Proc. ACM Interact. Mob. Wearable Ubiquitous Technol.},
	author = {Wang, Chao and Chu, Derck and Martens, Marieke},
	month = mar,
	year = {2025},
	pages = {20:1--20:27},
}

@inproceedings{goodge_can_2024,
	address = {New York, NY, USA},
	series = {{CHI} '24},
	title = {Can {You} {Hazard} a {Guess}?: {Evaluating} the {Effect} of {Augmented} {Reality} {Cues} on {Driver} {Hazard} {Prediction}},
	isbn = {979-8-4007-0330-0},
	shorttitle = {Can {You} {Hazard} a {Guess}?},
	url = {https://dl.acm.org/doi/10.1145/3613904.3642300},
	doi = {10.1145/3613904.3642300},
	urldate = {2025-11-04},
	booktitle = {Proceedings of the 2024 {CHI} {Conference} on {Human} {Factors} in {Computing} {Systems}},
	publisher = {Association for Computing Machinery},
	author = {Goodge, Thomas Alexander and Pollick, Frank and Brewster, Stephen Anthony},
	month = may,
	year = {2024},
	pages = {1--28},
}

@inproceedings{he_pedalling_2024,
	address = {New York, NY, USA},
	series = {{DIS} '24 {Companion}},
	title = {Pedalling into the {Future}: {Towards} {Enhancing} {Cycling} {Experience} {Using} {Augmented} {Reality}},
	isbn = {979-8-4007-0632-5},
	shorttitle = {Pedalling into the {Future}},
	url = {https://dl.acm.org/doi/10.1145/3656156.3663699},
	doi = {10.1145/3656156.3663699},
	urldate = {2025-11-04},
	booktitle = {Companion {Publication} of the 2024 {ACM} {Designing} {Interactive} {Systems} {Conference}},
	publisher = {Association for Computing Machinery},
	author = {He, Linjia and Wang, Hongyue and Goodwin, Sarah and Tag, Benjamin and Elvitigala, Don Samitha},
	month = jul,
	year = {2024},
	pages = {190--194},
}

@inproceedings{huang_safety_2025,
	address = {New York, NY, USA},
	series = {{CHI} {EA} '25},
	title = {Safety {Sense}: {Enhancing} {Urban} {Cyclists}' {Perceived} {Safety} during {Motorcycle} {Overtaking} through {Augmented} {Reality} {Warning} {System}},
	isbn = {979-8-4007-1395-8},
	shorttitle = {Safety {Sense}},
	url = {https://dl.acm.org/doi/10.1145/3706599.3719280},
	doi = {10.1145/3706599.3719280},
	urldate = {2025-11-04},
	booktitle = {Proceedings of the {Extended} {Abstracts} of the {CHI} {Conference} on {Human} {Factors} in {Computing} {Systems}},
	publisher = {Association for Computing Machinery},
	author = {Huang, Yu-Ti},
	month = apr,
	year = {2025},
	pages = {1--7},
}

@inproceedings{von_sawitzky_how_2024,
	address = {New York, NY, USA},
	series = {{AutomotiveUI} '24 {Adjunct}},
	title = {How to {Indicate} {Multiple} {Potential} {Hazards} to {Cyclists} on {Smart} {Glasses}? {Findings} from a {Focus} {Group} {Discussion} with {Research} {Experts} and {UX} {Students}},
	isbn = {979-8-4007-0520-5},
	shorttitle = {How to {Indicate} {Multiple} {Potential} {Hazards} to {Cyclists} on {Smart} {Glasses}?},
	url = {https://dl.acm.org/doi/10.1145/3641308.3685044},
	doi = {10.1145/3641308.3685044},
	urldate = {2025-11-04},
	booktitle = {Adjunct {Proceedings} of the 16th {International} {Conference} on {Automotive} {User} {Interfaces} and {Interactive} {Vehicular} {Applications}},
	publisher = {Association for Computing Machinery},
	author = {von Sawitzky, Tamara and Wintersberger, Philipp and Grauschopf, Thomas and Riener, Andreas},
	month = sep,
	year = {2024},
	pages = {178--183},
}

@inproceedings{wozniak_soil_2017,
	address = {New York, NY, USA},
	series = {{DIS} '17},
	title = {Soil, {Rock}, and {Snow}: {On} {Designing} for {Information} {Sharing} in {Outdoor} {Sports}},
	isbn = {978-1-4503-4922-2},
	shorttitle = {Soil, {Rock}, and {Snow}},
	url = {https://dl.acm.org/doi/10.1145/3064663.3064741},
	doi = {10.1145/3064663.3064741},
	urldate = {2025-11-05},
	booktitle = {Proceedings of the 2017 {Conference} on {Designing} {Interactive} {Systems}},
	publisher = {Association for Computing Machinery},
	author = {Wozniak, Paweł W. and Fedosov, Anton and Mencarini, Eleonora and Knaving, Kristina},
	month = jun,
	year = {2017},
	pages = {611--623},
}

@inproceedings{fedosov_towards_2016,
	address = {New York, NY, USA},
	series = {{UbiComp} '16},
	title = {Towards understanding digital sharing practices in outdoor sports},
	isbn = {978-1-4503-4462-3},
	url = {https://dl.acm.org/doi/10.1145/2968219.2968537},
	doi = {10.1145/2968219.2968537},
	urldate = {2025-11-05},
	booktitle = {Proceedings of the 2016 {ACM} {International} {Joint} {Conference} on {Pervasive} and {Ubiquitous} {Computing}: {Adjunct}},
	publisher = {Association for Computing Machinery},
	author = {Fedosov, Anton and Mencarini, Eleonora and Woźniak, Paweł and Knaving, Krisitna and Langheinrich, Marc},
	month = sep,
	year = {2016},
	pages = {861--866},
}

@inproceedings{niforatos_s-helmet_2016,
	address = {New York, NY, USA},
	series = {{AH} '16},
	title = {s-{Helmet}: {A} {Ski} {Helmet} for {Augmenting} {Peripheral} {Perception}},
	isbn = {978-1-4503-3680-2},
	shorttitle = {s-{Helmet}},
	url = {https://dl.acm.org/doi/10.1145/2875194.2875233},
	doi = {10.1145/2875194.2875233},
	urldate = {2025-11-05},
	booktitle = {Proceedings of the 7th {Augmented} {Human} {International} {Conference} 2016},
	publisher = {Association for Computing Machinery},
	author = {Niforatos, Evangelos and Elhart, Ivan and Fedosov, Anton and Langheinrich, Marc},
	month = feb,
	year = {2016},
	pages = {1--2},
}

@inproceedings{mencarini_becoming_2023,
	address = {New York, NY, USA},
	series = {{CHI} '23},
	title = {Becoming a {Speleologist}: {Design} {Implications} for {Coordination} in {Wild} {Outdoor} {Environments}},
	isbn = {978-1-4503-9421-5},
	shorttitle = {Becoming a {Speleologist}},
	url = {https://dl.acm.org/doi/10.1145/3544548.3581545},
	doi = {10.1145/3544548.3581545},
	urldate = {2025-11-05},
	booktitle = {Proceedings of the 2023 {CHI} {Conference} on {Human} {Factors} in {Computing} {Systems}},
	publisher = {Association for Computing Machinery},
	author = {Mencarini, Eleonora and Zambon, Tommaso},
	month = apr,
	year = {2023},
	pages = {1--12},
}

@inproceedings{barz_multisensor-pipeline_2021,
	address = {Montreal QC Canada},
	title = {Multisensor-{Pipeline}: {A} {Lightweight}, {Flexible}, and {Extensible} {Framework} for {Building} {Multimodal}-{Multisensor} {Interfaces}},
	isbn = {978-1-4503-8471-1},
	shorttitle = {Multisensor-{Pipeline}},
	url = {https://dl.acm.org/doi/10.1145/3461615.3485432},
	doi = {10.1145/3461615.3485432},
	language = {en},
	urldate = {2022-11-10},
	booktitle = {Companion {Publication} of the 2021 {International} {Conference} on {Multimodal} {Interaction}},
	publisher = {ACM},
	author = {Barz, Michael and Bhatti, Omair Shahzad and Lüers, Bengt and Prange, Alexander and Sonntag, Daniel},
	month = oct,
	year = {2021},
	pages = {13--18},
}

@inproceedings{bittner_interactive_2025,
	address = {Cagliari Italy},
	title = {Interactive {Multimodal} {Photobook} {Co}-{Creation} in {Virtual} {Reality}},
	isbn = {979-8-4007-1409-2},
	url = {https://dl.acm.org/doi/10.1145/3708557.3716355},
	doi = {10.1145/3708557.3716355},
	language = {en},
	urldate = {2025-09-16},
	booktitle = {Companion {Proceedings} of the 30th {International} {Conference} on {Intelligent} {User} {Interfaces}},
	publisher = {ACM},
	author = {Bittner, Sara-Jane and Leist, Robert Andreas and Kopácsi, László and Bhatti, Omair Shahzad and Mohamed Selim, Abdulrahman and Barz, Michael and Sonntag, Daniel},
	month = mar,
	year = {2025},
	pages = {146--151},
}

@article{wang_inattentional_2022,
    author = {Yuwei Wang and Yimin Wu and Cheng Chen and Bohan Wu and Shu Ma and Duming Wang and Hongting Li and Zhen Yang},
    title = {Inattentional Blindness in Augmented Reality Head-Up Display-Assisted Driving},
    journal = {International Journal of Human–Computer Interaction},
    volume = {38},
    number = {9},
    pages = {837--850},
    year = {2022},
    publisher = {Taylor \& Francis},
    doi = {10.1080/10447318.2021.1970434},
    URL = {https://doi.org/10.1080/10447318.2021.1970434},
    eprint = {https://doi.org/10.1080/10447318.2021.1970434}
}

@book{oviatt_handbook_2019,
editor = {Oviatt, Sharon and Schuller, Bj\"{o}rn and Cohen, Philip R. and Sonntag, Daniel and Potamianos, Gerasimos and Kr\"{u}ger, Antonio},
title = {The Handbook of Multimodal-Multisensor Interfaces: Language Processing, Software, Commercialization, and Emerging Directions},
year = {2019},
isbn = {9781970001754},
publisher = {Association for Computing Machinery and Morgan \& Claypool}
}

@online{Outdooractive2025_ATHM,
  author       = {{Outdooractive AG}},
  title        = {The ATHM},
  year         = {2025},
  note         = {Accessed: 2025‑11‑12},
  institution  = {Outdooractive},
  url          = {https://www.outdooractive.com/en/knowledgepage/the-athm/804017798/}
}


\end{document}